\pdfoutput=1

\documentclass[1p,preprint,authoryear]{elsarticle}

\makeatletter
\def\ps@pprintTitle{%
 \let\@oddhead\@empty
 \let\@evenhead\@empty
 \def\@oddfoot{}%
 \let\@evenfoot\@oddfoot}
\makeatother

\usepackage{a4wide} 
\usepackage{setspace} 
\usepackage{floatflt}
\usepackage{float} 
\usepackage{xspace}
\usepackage{pdfpages}
\usepackage{amsmath}
\usepackage{amssymb}
\usepackage{lineno}
\usepackage{url} 

\usepackage{booktabs}
\usepackage{siunitx} 

\usepackage{color}
\definecolor{RoyalBlue}{cmyk}{1, 0.50, 0, 0}

\usepackage[colorlinks]{hyperref}
\AtBeginDocument{%
  \hypersetup{
    allbordercolors={1 1 1},
    allcolors={RoyalBlue}}}

\begin{document}
\begin{frontmatter}

\indent \textcolor{RoyalBlue}{Cite as: Grohmann, C.H., 2018. Evaluation of TanDEM-X DEMs on selected Brazilian sites: comparison with SRTM, ASTER GDEM and ALOS AW3D30. \textit{Remote Sensing of Environment}. 212:121-133. doi:\href{http://dx.doi.org/10.1016/j.rse.2018.04.043}{10.1016/j.rse.2018.04.043}} \\

\vspace{15pt} 
\title{Evaluation of TanDEM-X DEMs on selected Brazilian sites: comparison with SRTM, ASTER GDEM and ALOS AW3D30}

\author{Carlos H. Grohmann}
\address{Institute of Energy and Environment, University of S\~{a}o Paulo, S\~{a}o Paulo, 05508-010, Brazil}
\ead{guano@usp.br}
\ead[url]{http://www.iee.usp.br, http://carlosgrohmann.com}

\begin{abstract}
A first assessment of the TanDEM-X DEMs over Brazilian territory is presented through a comparison with SRTM, ASTER GDEM and ALOS AW3D30 DEMs in seven study areas with distinct geomorphological contexts, vegetation coverage, and land use. Visual analysis and elevation histograms point to a finer effective spatial (i.e., horizontal) resolution of TanDEM-X compared to SRTM and ASTER GDEM. In areas of open vegetation, TanDEM-X lower elevations indicate a deeper penetration of the radar signal. DEMs of differences (DoDs) allowed the identification of issues inherent to the production methods of the analyzed DEMs, such as mast oscillations in SRTM data and mismatch between adjacent scenes in ASTER GDEM and ALOS AW3D30. A systematic difference in elevations between TanDEM-X 12~m, TanDEM-X 30~m, and SRTM was observed in the steep slopes of the coastal ranges, related to the moving-window process used to resample the 12~m data to a 30~m pixel size. It is strongly recommended to produce a DoD with SRTM before using ASTER GDEM or ALOS AW3D30 in any analysis, to evaluate if the area of interest is affected by these problems. The DoDs also highlighted changes in land use in the time span between the acquisition of SRTM (2000) and TanDEM-X (2013) data, whether by natural causes or by human interference in the environment. The results show a high level of detail and consistency for TanDEM-X data, indicate that the effective horizontal resolution of SRTM is coarser than the nominal 30~m, and highlight the errors in ASTER GDEM and ALOS AW3D30 due to mismatch between adjacent scenes in the photogrammetric process.
\end{abstract}


\end{frontmatter}

\defcitealias{ERSDAC2009}{ERSDAC, 2009}
\defcitealias{NGA1997}{NGA, 1997}
\defcitealias{NGA2014}{NGA, 2014}
\defcitealias{JPL2014}{JPL, 2014}
\defcitealias{JAXA2017}{JAXA, 2017}
\defcitealias{FGDC1998}{FGDC, 1998}

\section{Introduction}
\label{sec:intro}

Global Digital Elevation Models (DEMs) have become essential data for research in areas such as geomorphology, climatology, oceanography and biodiversity, with applications as diverse as the development of geopotential global models \citep{Arabelos2000}, evaluation of glacier volume change \citep{Berthier2006}, climatic modeling \citep{Moore1991,Thomas2004}, vegetation mapping \citep{Kellndorfer2004,OLoughlin2016} or navigation systems for commercial aviation \citep{Fox2008}.

Global or quasi-global DEMs currently available include SRTM \citep{Farr2007}, ASTER GDEM \citep{Tachikawa2011} and ALOS AW3D DEM \citep{Tadono2015}. SRTM is likely the most successful and widely used DEM to date, despite limitations such as presence of voids caused by radar shadowing and lack of coverage at high latitudes. ASTER GDEM and ALOS AW3D are built based on photogrammetric processing of optical satellite imagery, thus containing artifacts and voids due to cloud cover in the original images. 

TanDEM-X DEM is a new dataset produced by the German Aerospace Center (DLR) with global coverage, spatial (i.e., horizontal) resolution of 12~m, and is expected to represent a new standard in global DEMs regarding geometric resolution, accuracy and ability to depict complex topography \citep{Krieger2007,Zink2014,Rizzoli2017}.

Early and intermediate products of the TanDEM-X mission have been evaluated for their height accuracy by \cite{Gruber2012,Wessel2014,Wecklich2015,Baade2016} and \cite{Wecklich2017}, while \cite{Erasmi2014} assessed its applications in archeology, and \cite{Pandey2013} conducted a comparison with SRTM in the Himalayas.

In this paper, a first assessment of the final TanDEM-X DEMs over Brazilian territory is presented through a comparison with SRTM, ASTER GDEM and ALOS AW3D30 DEMs in seven study areas with distinct geomorphological contexts, vegetation coverage and land use. The results show a high level of detail and consistency for TanDEM-X data, indicate that the effective spatial resolution of SRTM is coarser than the nominal 30~m, and highlight the errors in ASTER GDEM and ALOS AW3D30 due to mismatch between adjacent scenes in the photogrammetric process. Additionally, DEMs of differences proved to be a simple and effective tool to perform a preliminary evaluation of a DEM, and are recommended prior to any analysis which intents to use ASTER GDEM or ALOS AW3D DEMs.

\begin{figure*}[!ht]
    \centering
    \includegraphics[width=0.8\textwidth]{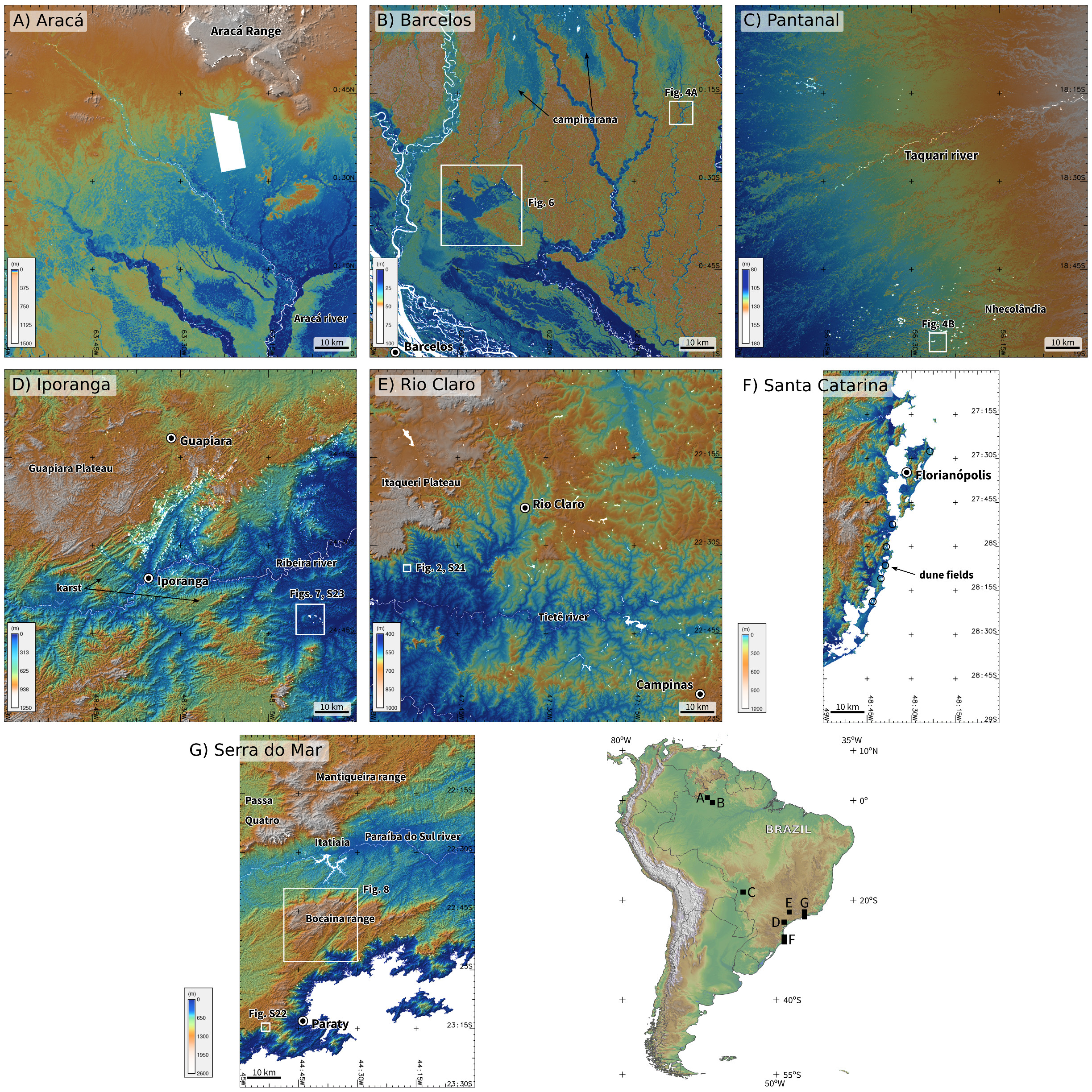} 
    \caption{Shaded relief maps of studied areas with indication of places or landscape features referred in the text (TanDEM-X 30~m data, shaded relief illumination from \ang{315}N, \ang{20} above horizon).}
    \label{fig:areas}
\end{figure*}

\subsection{Study areas}
\label{sec:study_areas}

Seven areas were selected for analysis, in order to represent a wide range of geomorphological contexts, vegetation coverage and land use. Each area is defined by one or two $\ang{1}\times\ang{1}$ tiles of the analyzed DEMs. Figure \ref{fig:areas} shows shaded relief images of TanDEM-X (30~m resolution) for all areas; satellite imagery is presented in Supplemental Figure S1. In this paper, areas are referred to after geomorphological features, cities or location; labels in Figure \ref{fig:areas} are provided as guides to the reader and do not represent an exhaustive description of all landscape elements.

\begin{figure*}[!hbt]
    \centering
    \includegraphics[width=0.8\textwidth]{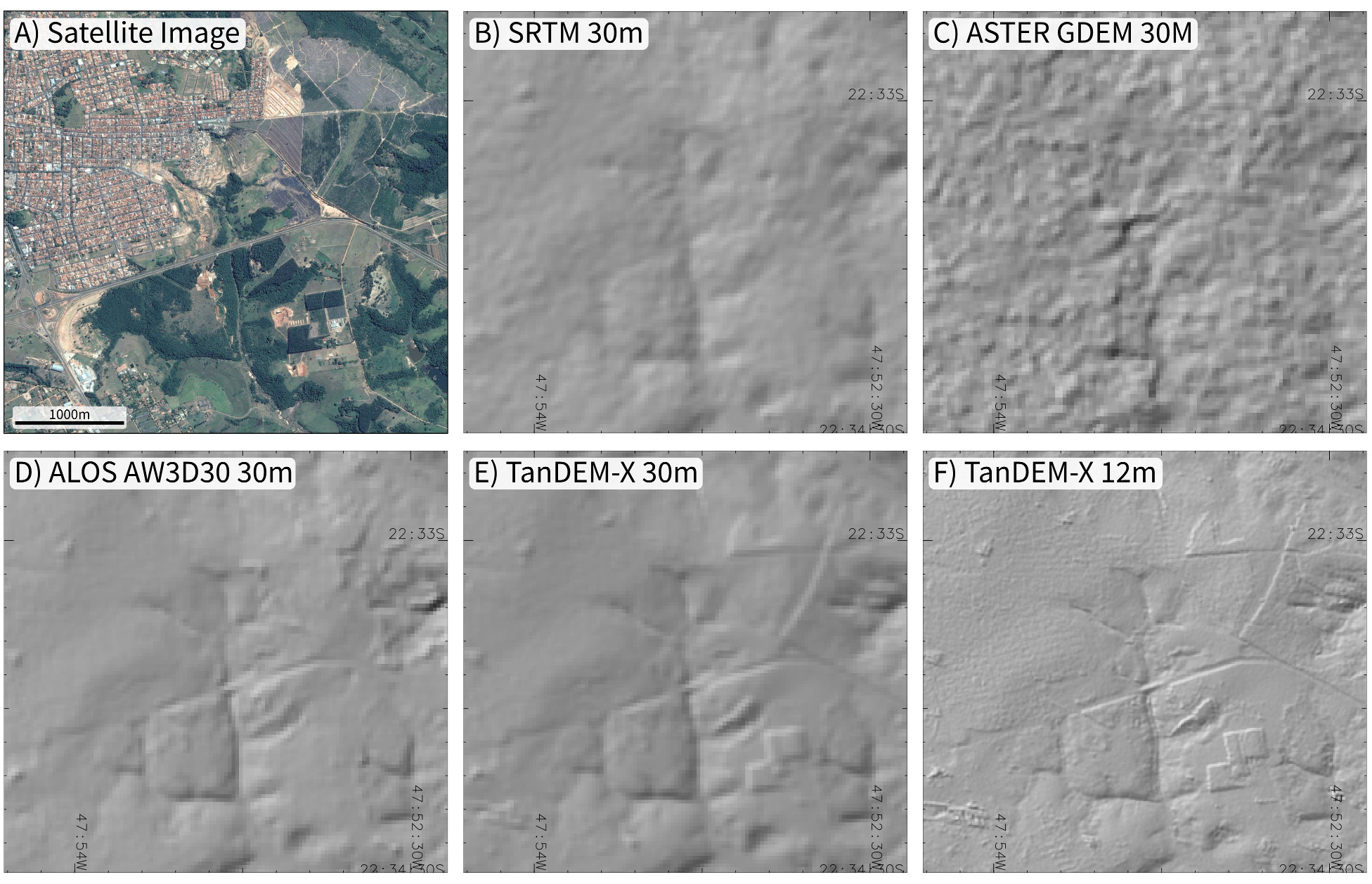} 
    \caption{Inset of the Rio Claro area (see location in Fig.\ref{fig:areas}E), where the level of detail depicted by each DEM is show by shaded relief maps. All shaded relief images have illumination from \ang{315}N, \ang{20} above horizon. A) Satellite image (image date: 04-18-2016); B) SRTM (30~m); C) ASTER GDEM (30~m); D) ALOS AW3D30 (30~m); E) TanDEM-X (30~m); F) TanDEM-X (12~m). Satellite imagery \textcopyright CNES/Airbus, powered by Google.}
    \label{fig:shades_tucum}
\end{figure*}

Shaded relief images for all DEMs and areas are shown in the Supplemental Figures S2--S8; it is possible to observe the large area of voids (no data) in ALOS AW3D30 data, likely due to cloud coverage, notably in the Arac\'a, Barcelos and Iporanga areas. Voids in TanDEM-X DEMs are not present in the original data, and result from a filtering process based on the Water Indication Mask \citep[WAM --][]{Wendleder2013}  supplementary information layer (see Sec.~\ref{sec:wam}). An inset of the Rio Claro area (Fig.~\ref{fig:shades_tucum}) highlights the level of detail resolved by TanDEM-X data. Located in the outskirts of the small town of S\~ao Pedro, it shows suburban and rural lands with a large linear erosional feature in its central zone. At resolution of 30~m, the erosion is barely seen in SRTM and ASTER GDEM, while it can be identified in ALOS AW3D30 and TanDEM-X 30~m. With 12~m resolution, it is possible to delineate not only the erosion gully, but also streets, roads and agricultural areas.

Two areas are located within the Amazon Forest in northern Brazil: the Arac\'a range area (tile N00W064 -- Fig.~\ref{fig:areas}A) and the Barcelos city and Negro River floodplain area (tile S01W063 -- Fig.~\ref{fig:areas}B). These two areas are dominated by a smooth and low-lying topography generally below $200$~m altitude, with wide meandering belts of seasonally-flooded rivers. Locally, the evergreen Amazon Rain Forest is replaced by naturally occurring areas of white sands soils which sustain a savanna vegetation called campinaranas.  The monotony of the landscape is disturbed by the Arac\'a range, an isolated plateau (or Tepui) with elevations above $1\,000$~m in altitude and scrub vegetation. 

In central-western Brazil, the Pantanal wetlands area (tile S19W057 -- Fig.~\ref{fig:areas}C) has a smooth and flat topography characterized by permanently flooded areas and a drainage network running through narrow alluvial plains subject to seasonal floods. In its southeast, a region known as Nhecol\^andia shows an unique landscape, with thousands of small saline lakes. 

Located in inland S\~ao Paulo State in southeastern Brazil, Iporanga (tile S25W049 -- Fig.~\ref{fig:areas}D) and Rio Claro (tile S23W048 -- Fig.~\ref{fig:areas}E) areas represent landscapes of fluvial dissection with mixed topography, including rolling hills, erosion escarpments, plateaus, elongated ridges and local karstic features. Land cover includes agriculture, pasture and large urban areas in the Rio Claro tile while the Iporanga tile also include a wide area of evergreen Atlantic Forest, preserved in a series of conservation units. 

Santa Catarina (tiles S28W049, S29W049 -- Fig.~\ref{fig:areas}F) and Serra do Mar (tiles S23W045, S24W045 -- Fig.~\ref{fig:areas}G) areas represent the Atlantic Coastal Range of Brazil, with steep scarps and altitudes ranging from sea level up to $2\,800$~m at the alkaline massifs of Passa Quatro and Itatiaia in the Mantiqueira Range (Serra do Mar area). Atlantic Forest vegetation is more preserved in the steep slopes of the ranges and in conservation units such as the Itatiaia National Park. In Santa Catarina, small dune fields developed on the coastal plain (marked with circles in Fig.~\ref{fig:areas}F).

\subsection{Datasets descriptions}
\label{sec:dems_description}

This section presents a brief descriptions of the analyzed datasets. The recently released Multi-Error-Removed Improved-Terrain DEM \citep[MERIT DEM --][]{Yamazaki2017} was not included in this study due its coarser spatial resolution (03'').

Table \ref{tbl:dems_accuracy} summarizes the characteristics of imaging systems and accuracy of the datasets used in this study. Horizontal accuracy can be expressed as RMSE or as circular error at 90-95\% confidence level (CE90, CE95) and vertical accuracy as RMSE or as linear error at 90-95\% confidence level (LE90, LE95). 

It is worth noting that the DEMs selected for this analysis could also be referred to as DSMs (Digital Surface Models), in the sense that they do not represent the 'bare' topographic surface in vegetated or urban areas (in which case they would be called Digital Terrain Models - DTMs), due the weak penetration of the radar signal in dense vegetation for TanDEM-X and SRTM and the use of optical images in ASTER GDEM and ALOS AW3D30, which are also sensitive to cloud coverage. In the case of SRTM, the C band data has been shown to penetrated significantly into the vegetation canopy \citep{Carabajal2006,Hofton2006}.

\subsubsection{TanDEM-X}
\label{sec:tandemx}

The TanDEM-X mission (TerraSAR-X add-on for Digital Elevation Measurements) goal was to produce a Global DEM with 12~m spatial resolution, from radar interferometry (InSAR) with two satellites (TerraSAR-X and TanDEM-X) in a controlled orbit with a baseline of 250-500~m \citep{Krieger2007,Martone2012,Gruber2016,Rizzoli2017}. The configuration of the sensors and orbit allowed across-track and along-track interferometry. The project accomplished complete imaging of Earth at least twice, with additional coverage in areas of complex topography, including the repositioning of the orbits to avoid radar shadowing in mountainous terrains \citep{Brautigam2013,Zink2014,Wecklich2017}. 

The instruments on both satellites are X-band synthetic aperture radars with 3.1~cm wavelength and full polarization capability \citep{Krieger2013}. The absolute accuracies (i.e., uncertainties with respect to a horizontal datum or reference height) requirement for the TanDEM-X product is $<10$~m horizontal and vertical, while the relative vertical accuracy (uncertainty in height between two points in a $\ang{1}\times\ang{1}$ area) is $2$~m for areas with slope $\leq20\%$ and $4$~m when slope $>20\%$ \citep{Wessel2016}. 

WorldDEM is the commercial product of the TanDEM-X Mission, realized as a Public Private Partnership between the German Aerospace Center (Deustches Zentrum f\"ur Luft- und Raumfahrt -- DLR) and Airbus Defence and Space. DLR is responsible for providing TanDEM-X data to the scientific community. The reader is referred to \cite{Krieger2007,Gonzalez2010,Rizzoli2012} and \cite{Rizzoli2017} for a detailed description of the TanDEM-X mission and global DEM generation.

\subsubsection{SRTM}
\label{sec:srtm}

The Shuttle Radar Topography Mission was a cooperation among NASA, the U.S. National Geospatial-Intelligence Agency (NGA), U.S. Department of Defense (DoD), DLR, and the Agenzia Spaziale Italiana (ASI, Italy). The STS-99 space mission of the Endeavour Space Shuttle flew during 11 days in February 2000; its main objective was the topographical mapping of continental areas between \ang{60}N and \ang{60}S (about 80\% of the Earth's land masses) with InSAR \citep{Farr2000,Zyl2001,Rabus2003}. A detailed review of the SRTM mission is given by \cite{Farr2007}.

Two synthetic aperture radars operated during the SRTM mission: a C band system (5.6 cm, SIR-C) and an X band system (3.1 cm, X-SAR). While the C radar generated a contiguous mapping coverage, the X radar generated data along discrete swaths 50 km wide \citep{Farr2007}. SRTM data used in this study is from the C band system. The accuracy requirements of the mission, $<20$~m of geolocation error and $<16$~m of vertical error, were exceeded by a factor of almost 2, with errors $<12.6$~m in geolocation and $<9$~m vertical in the final product \citep{Rodriguez2005}.

Three official versions of SRTM have been released. The goal of the last official version (V3 or ``SRTM Plus'') was the complete elimination of voids, which were filled mainly with data from ASTER GDEM. In 2014, a global SRTM V3 data with 01'' resolution (${\sim}$30 m at the Equator) was publicly released \citepalias{NGA2014, JPL2014}. The next version of a global DEM based on a full reprocessing of SRTM radar data, NASADEM, is under development \citep{Crippen2016, Simard2016}.

\subsubsection{ASTER GDEM}
\label{sec:astergdem}

The ASTER sensor \citep[Advanced Spaceborne Thermal Emission and Reflection Radiometer --][]{Yamaguchi1998} was launched in December 1999 onboard the Terra satellite, with the capability of generating along-track stereoscopic images on the Near Infra-Red wavelength (0.78--\SI{0.86}{\micro\meter}) with telescopes aligned to nadir (3N band) and backwards (3B band), with 15~m spatial resolution. 

In 2009 ASTER GDEM (Global DEM) version 1 was released, covering all land areas between \ang{83}N and \ang{83}S \citepalias{ERSDAC2009}. ASTER GDEM V.1 was produced by automatically processing the entire ASTER archive (about $1\,500\,000$ scenes acquired from 2000 to 2008) \citep{Abrams2010, Tachikawa2011}. ASTER GDEM V.2 was released in 2011 \citep{Tachikawa2011} and improves on the first version on the processing algorithms, inclusion of scenes acquired between 2008 and 2011 (about $250\,000$ scenes), better georreferencing of data and increase of effective spatial resolution from 120~m to 70~m. At 95\% confidence ASTER GDEM has an estimated accuracy of 30~m horizontal and 20~m vertical \citep{Tachikawa2011a}.

\subsubsection{ALOS WORLD 3D (AW3D)}
\label{sec:alos3d}

The PRISM sensor \citep[Panchromatic Remote-Sensing Instrument for Stereo Mapping --][]{Tadono2009, Shimada2010} was launched in January 2006 onboard the ALOS (Advanced Land Observing Satellite) satellite, with the capability of generating a triplet of along-track stereoscopic Panchromatic (0.52--\SI{0.77}{\micro\meter}) images at nadir (NDR), forward (FWD) and backwards (BWD), with 2.5~m resolution. 

In its five years of operation, ALOS produced approximately 6.5 million scenes covering the entire globe. An automated process of all scenes with less than 30\% cloud cover (about 3 million scenes) was used to generate a global DEM with 5~m resolution (0.15'') \citep{Tadono2014,Takaku2014,Tadono2015}. Although the 5~m dataset is distributed commercially, a 30~m resolution version (AW3D30) is freely available and was used in this study. Accuracies are reported only for the 5~m dataset as an RMSE of 5~m for horizontal and vertical \citep{Takaku2014}.

\begin{table*}[!hbt]
    \caption{Characteristics and global accuracy of the datasets used in this study. }
    \label{tbl:dems_accuracy}
    \begin{center}
    \resizebox{0.85\textwidth}{!} {
    \begin{tabular}{llllll}
    \toprule
    Dataset      & Imaging System  & Wavelenght                     & Pixel Spacing & Horizontal Accuracy & Vertical Accuracy \\
    \midrule         
    TanDEM-X     & SAR X band    & 3.1 cm                         & 12~m, 30~m    & $<10$~m (CE90)      & $<10$~m (LE90)  \\
    SRTM         & SAR C band    & 5.66 cm                        & 30~m          & $<20$~m (CE90)      & $<16$~m (LE90)  \\
    ASTER GDEM   & Optical         & 0.78--\SI{0.86}{\micro\meter}  & 30~m          & $30$~m (CE95)       & $20$~m (LE95)  \\
    ALOS AW3D    & Optical         & 0.52--\SI{0.77}{\micro\meter}  & 5~m           & $5$~m (RMSE)        & $5$~m (RMSE)  \\
    \bottomrule
    \end{tabular}
    } 
    \end{center}
\end{table*}

\subsection{Software and Data}
\label{sec:soft_data}

In order to streamline the process and ensure reproducibility  \citep{Barnes2010}, data processing was performed in GRASS-GIS  version 7.2 \citep{Neteler2012, GRASS2017} through Python scripts \citep{Python2013} using the Pygrass library  \citep{Zambelli2013} to access GRASS’ datasets. Statistical analyses were performed with the Python libraries Scipy, Numpy, Pandas, Seaborn and Matplotlib \citep{Oliphant2006, Hunter2007, McKinney2011, SciPy2013, seaborn2016}. The scripts and associated data files are available on GitHub at \url{https://git.io/vQTyp}.

TanDEM-X data for this work were provided by DLR, with spatial resolutions of 12~m (0.4 arcsec) and 30~m (1 arcsec); the 30~m version is generated from the unweighted mean values of the underlying 12~m pixels \citep{Wessel2016}. SRTM V3 (30~m resolution) and ASTER GDEM V2 (30~m resolution) data were downloaded from the NASA EOSDIS Land Processes Distributed Active Archive Center (LP DAAC -- \url{https://lpdaac.usgs.gov/}). ALOS AW3D30 data (30~m resolution) were downloaded from the JAXA Earth Observation Research Center (\url{http://www.eorc.jaxa.jp/ALOS/en/aw3d30/index.htm}).

\subsection{Water masking}
\label{sec:wam}

The TanDEM-X Water Indication Mask (WAM) is provided as an auxiliary file which can be used in the DEM editing process. As water surfaces usually show lower coherence in interferometric data due to temporal de-correlation and low backscattering, the corresponding elevation values derived are random and produce a noisy surface \citep{Wendleder2013}. The values in WAM are coded in a bit mask, where each bit value reflects the number of acquisitions with detected water by thresholds of the SAR amplitude or coherence. Islands smaller than 1~ha ($100\times100$~m$^2$) and water bodies smaller than 2~ha ($200\times100$~m$^2$) are not included in WAM. 

\begin{figure*}[!hbt]
    \centering
    \includegraphics[width=0.8\textwidth]{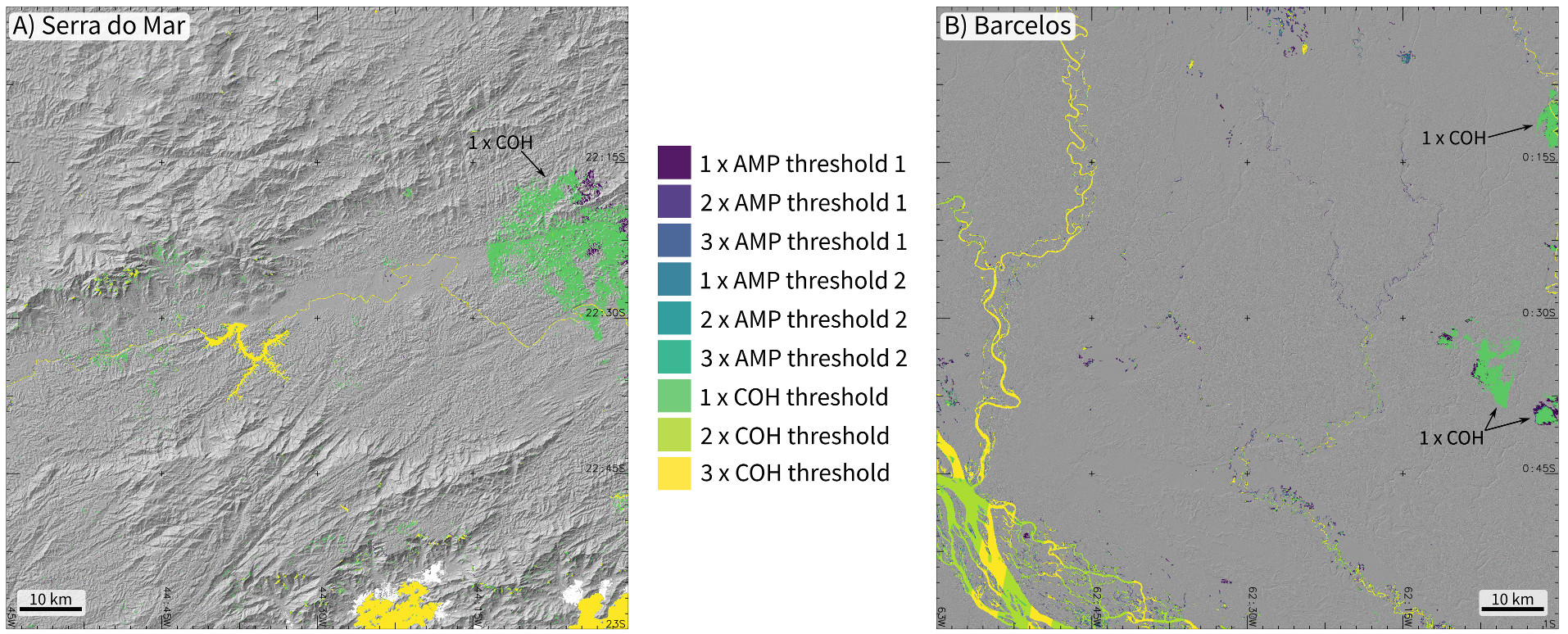} 
    \caption{Classes of the Water Indication Mask (WAM) over shaded relief images. Note the areas indicated as `1xCOH', where the coherence threshold was flagged only one time, and did not corresponded to water bodies.}
    \label{fig:water_mask}
\end{figure*}

\cite{Wendleder2013} showed that water bodies derived from coherence thresholds have a higher level of correctness when compared to a reference dataset than those derived from amplitude thresholds, although less rich in detail. \cite{Wessel2016} details the bit mask in WAM: selecying byte values from 3 to 32 results in a water mask based on the amplitude only, while values from 33 to 127 refer to the coherence thresholds. After an initial analysis of WAM, byte values $>$65 were selected as a water mask and matching pixels in the TanDEM-X DEMs were marked as non-valid. Values between 33 and 64 indicate areas where the coherence threshold was flagged only one time, and did not corresponded to water bodies (Fig.\ref{fig:water_mask}).

\subsection{Orthometric height conversion}
\label{sec:heightconversion}

TanDEM-X elevations correspond to ellipsoidal heights referenced to the WGS84-G1150 ellipsoid \citep{Gruber2012}, while values in SRTM, ASTER GDEM and ALOS AW3D30 are altitudes (orthometric heights) referenced to the EGM96 geoid \citep{Lemoine1998,Farr2007,Tachikawa2011}\citepalias{JAXA2017}. 

Ellipsoidal height $h$ can be obtained by adding the geoid undulation or separation $N$ to the orthometric height $H$ ($h=H+N$) \citep{Johnson2014,vanSickle2008}. In the approach used to convert SRTM original ellipsoidal heights to orthometric, N was calculated by evaluating EGM96 at 0.1 degree intervals using a full 360$\times$360 harmonic expansion, followed by bilinear interpolation \citep{Farr2007}. Following \cite{Baade2016} and \cite{Hu2017}, the SRTM process to determine geoid undulation was assumed to be applicable to ASTER GDEM and ALOS AW3D30. In this paper, EGM96 was evaluated for each study area at 0.1 degree intervals using the \texttt{F477.F} program provided by NGA at \url{http://earth-info.nga.mil/GandG/wgs84/gravitymod/egm96/egm96.html} and bilinear interpolation was used to create a raster with 30~m resolution. Using raster algebra, the EGM96 raster was added to SRTM, ASTER GDEM and ALOS AW3D30 DEMs, converting the elevation from orthometric heights to ellipsoidal heights referenced to WGS84. The changes in elevation after height conversion can be seen in the descriptive statistics presented in Supplemental Tables S1--S7.

\subsection{Comparison of datasets}
\label{sec:comparison_methods}

Comparison of the datasets was carried out based on: 1) descriptive statistics of the DEMs; 2) descriptive statistics of slope; 3) analysis of contour lines and 4) analysis of the differences in elevation between TanDEM-X 12~m and each DEM.

Descriptive statistics are presented as histograms of elevation values and as tables with values of minimum, maximum, range of values, mean (average), median, standard deviation, skewness, kurtosis, 25\textsuperscript{th} and 75\textsuperscript{th} percentiles. Slope is discussed in terms of its descriptive statistics, histograms, plots of mean slope per elevation and slope maps, useful to visualize local changes in elevation due to noise or artifacts in the DEMs. Contour lines were compared about their length and number of lines derived from each DEM. Differences in the elevations of DEMs are analyzed according to error metrics, maps and histograms. 

The vertical accuracy of DEMs is usually computed from the differences between the dataset being analyzed and co-located values from an independent source of higher accuracy \citep{Willmott2005,Wechsler2007,Hebeler2009,Reuter2009,Baade2016} although metrics derived from a set of points that are not evenly distributed in the study area (such as dGPS tracks or geodesic stations) might not address the spatial variation of errors \citep{Unwin1995}, and may fail the normality requirement for RMSE \citep{Monckton1994}.

In the absence of ground control data, a higher resolution surface can be considered as reference of 'true' elevation. LiDAR has been used in the evaluation of DEMs generated with Structure-from-Motion/Multi-View-Stereo (SfM/MVS) \citep{James2012,Westoby2012,Fonstad2013,Clapuyt2016,Casagli2017,Cook2017} and also in assessments of SRTM and ASTER GDEM \citep{Gonga-Saholiariliva2011,Grohmann2013gmorph,DeWitt2015}.

In this study TanDEM-X 12~m was considered the reference for analysis of differences between DEMs. 

Maps of the differences between DEMs (here called DEMs of differences -- DoDs) can be used to determine, for example, areas with changes in the land surface \citep{James2012,Westoby2012,Clapuyt2016}, map vegetation height \citep{Carabajal2006,OLoughlin2016,Grohmann2015poznan} or to highlight the spatial pattern of systematic errors, such as misregistration between datasets \citep{Rodriguez2006,Niel2008} or dome-shaped distortions in DEMs calculated with SfM/MVS \citep{Wackrow2011,James2014}.

DoDs were calculated with raster algebra \citep{Shapiro1991} by subtracting the elevations of each DEM from TanDEM-X 12~m. Positive values represent areas where TanDEM-X has higher elevations than the other DEM, and vice-versa. To avoid artifacts in the resulting maps, SRTM, TanDEM-X 30~m, ASTER GDEM and ALOS AW3D30 were resampled to 12~m with bicubic interpolation beforehand. 

The vertical Root Mean Square Error (RMSE) is a common metric used extensively in the Geosciences to measure the accuracy of continuous variables, such as elevation of DEMs \citep[e.g., ][]{Nikolakopoulos2006,Willmott2006,Smith2015,Gesch2016,Satge2016} and is estimated by

\begin{equation}
    \label{eq:rmse}
    RMSE = \sqrt{\dfrac{1}{n}\textstyle\sum_{1}^{n}\Big[(z_{TDX_{12m}} - z_{DEM})^{2}\Big]}
\end{equation}

where $z_{TDX_{12m}}$ is the elevation of TanDEM-X 12~m, $z_{DEM}$ is the elevation for each DEM (resampled to 12~m) and \textit{n} is the total number of valid points. Missing observations due to voids in the DEMs were filtered pairwise.

The Standard Error \citep{Greenwalt1962,Congalton2008} is defined as

\begin{equation}
    \label{eq:stde}
    STDE = \sqrt{\dfrac{1}{n-1}\textstyle\sum_{1}^{n}(z_{TDX_{12m}} - \bar{z}_{DEM})^{2}}
\end{equation}

where $\bar{z}_{DEM}$ is the arthimetic mean of ${z}_{DEM}$. 

Assuming that the vertical errors are normally distributed, linear error is proportional to the Standard Error \citep{Greenwalt1962,FGDC1998,Maune2007a,Congalton2008}. At confidence levels of 90\%, 95\% and 99.73\% ($3\sigma$), it is calculated as

\begin{equation}
    \label{eq:le90}
    LE90 = 1.6449 \times STDE
\end{equation}

\begin{equation}
    \label{eq:le95}
    LE95 = 1.9000 \times STDE
\end{equation}

\begin{equation}
    \label{eq:le99}
    LE3\sigma = 3.0000 \times STDE
\end{equation}

\vskip 0.5cm
\section{Results and Discussions}
\label{sec:results}

\subsection{Descriptive statistics}
\label{sec:compare_dems}

Descriptive statistics for all DEMs and areas are presented in Supplemental Tables S1--S7, while the distribution of the elevation values can be seen as histograms in Supplemental Figure S9. 

In general, all histograms show similar curves for the analyzed DEMs, with local differences in the shape of the curve or in the position of peaks. 

In the Arac\'a area, all histograms are strongly asymmetrical, and only the 0--150~m interval is shown in Supplemental Figure S9A, with the full histogram (for SRTM) as an inset. In this elevation interval, SRTM and ASTER GDEM have unimodal distributions with peaks around 40--50~m. TanDEM-X data (12~m and 30~m) and ALOS AW3D30 have bimodal distributions with peaks at 20--30~m and 50~m. A similar situation occurs in the Barcelos area (Supplemental Figure S9B), where all DEMs show a major peak in the histogram at ${\sim}45$~m and secondary peaks at ${\sim}20$~m and ${\sim}30$~m. 

Visual inspection of satellite and shaded relief images reveal that large patches of campinaranas occur in the areas, which could account for the 20--30~m elevation peaks in the histograms, considering a finer effective spatial resolution of TanDEM-X than that of SRTM \citep[${\sim}$60~m --][]{Guth2006} or ASTER GDEM \citep[${\sim}$70-120~m --][]{Reuter2009c, Miliaresis2011,Tachikawa2011}, and deeper penetration of the radar signal in open vegetation. In the case of ALOS AW3D30, it is likely reflecting the 5~m pixel of the original AW3D DEM: the finer pixel size is capable of detecting open areas between trees, which can influence the averaging of the resampling process and result in a 30~m pixel with elevation value lower than the mean canopy height of its ground area.

In the Pantanal area, distribution of elevation has unimodal shape, minimum at ${\sim}$100~m, maximum at ${\sim}$150~m and peak around 110--120~m (Supplemental Figure S9C). TanDEM-X (12~m and 30~m) is ${\sim}$5~m lower in elevation than SRTM, ASTER GDEM and ALOS AW3D30 for the whole area, as it can be seen in the histograms, where the curves for TanDEM-X have similar shape, but are plotted to the left of SRTM, ASTER and ALOS data. 

Histograms for Rio Claro, Iporanga and Serra do Mar areas (Supplemental Figures S9D,E,G) are very similar for all datasets. These areas have moderate to high relief and mixed land cover, from smooth gentle hills in rural areas to forested mountainous terrains. The response of these land cover classes to optical sensors or C/X band radar is comparable (close to bare ground over agricultural or pasture areas and top of canopy for dense forests), resulting in similar DEMs, despite the variability in land cover and relief.

The elevation for the Santa Catarina area is highly asymmetrical, and in the same way as the Arac\'a area, only the 0--150~m interval is shown in Supplemental Figure S9F, with the full histogram for SRTM as an inset. The major differences between the datasets occur in the 0--40~m range, where TanDEM-X data show a peak at ${\sim}$2~m, SRTM and ALOS AW3D30 at ${\sim}$5~m and ASTER GDEM peak at ${\sim}$10~m. This reflects the geomorphology of the area, with a large coastal plain and presence of aeolian dune fields, which can show radar shadowing effects and can be difficult to model via photogrammetry due the lack of contrast in homogeneous sand landforms.

\subsection{Slope}
\label{sec:slope}

\begin{figure*}[!hbt]
    \centering
    \includegraphics[width=0.8\textwidth]{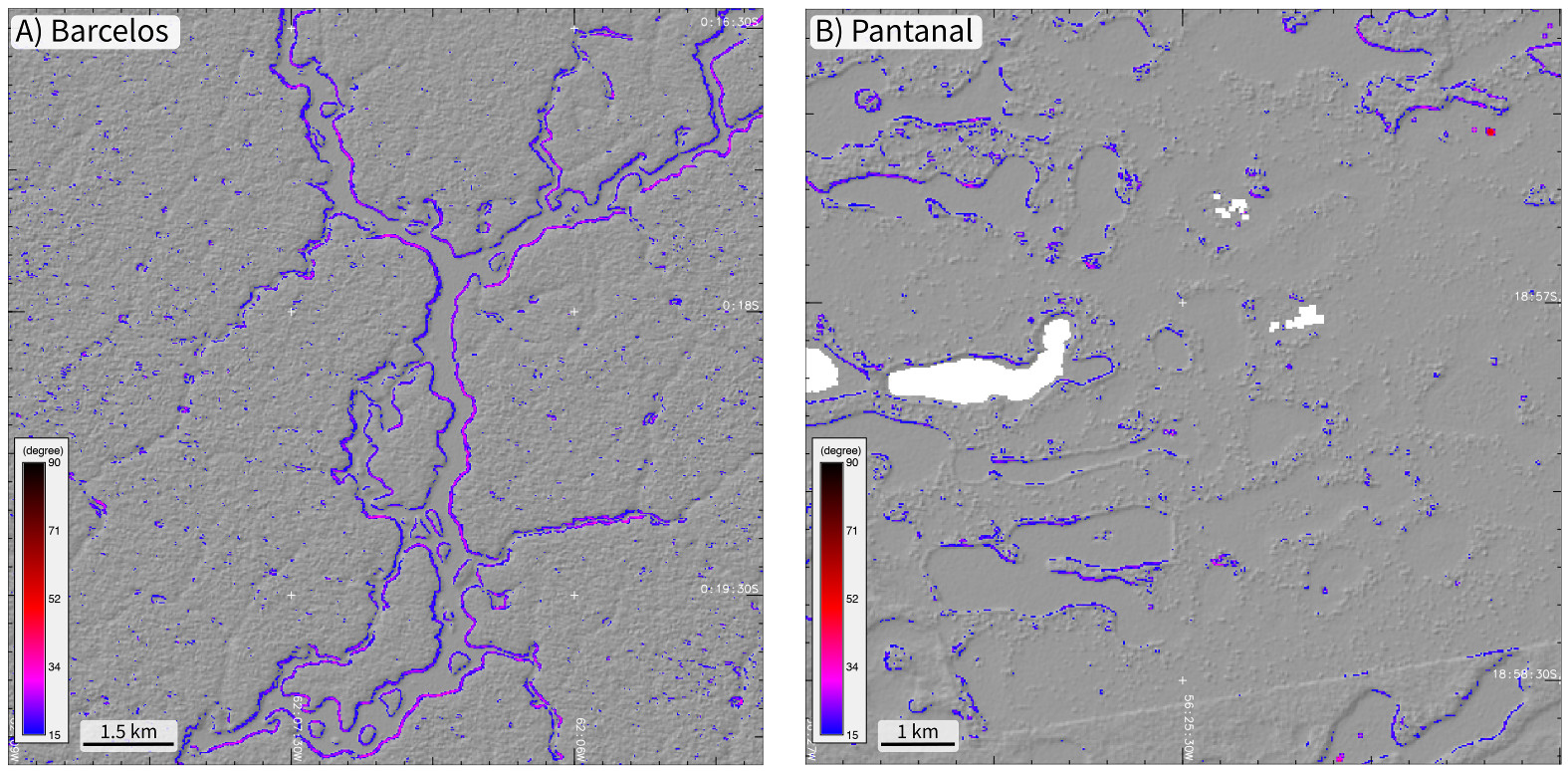} 
    \caption{Examples of areas with steep slope values corresponding to river terraces escarpments or borders of forested areas.}
    \label{fig:shade_high_slope}
\end{figure*}

Maps of slope for all study areas and DEMs are presented in Supplemental Figures S10--S16; summary statistics are presented in Supplemental Tables S8--S14, the distribution of slope values can be seen as histograms in Supplemental Figure S17, while Supplemental Figure S18 shows plots of mean slope per elevation \citep{Guth2006,Grohmann_2008ijgis}, which provide an effective way of comparing the response of each DEM to local variations of the topographic surface.

In general, there is a good agreement between slopes derived from all datasets, but TanDEM-X 12~m shows steeper slopes than the other DEMs due its finer spatial resolution, since slope tends to decrease as DEM resolution becomes coarser \citep{Chow2009,Chen2013,Grohmann2015cageo}.   

In smooth and flat areas with a landscape dominated by fluvial processes, such as Barcelos and Pantanal, spikes in TanDEM-X 12~m mean slope values that differ substantially from the values of the other DEMs can be interpreted as river terraces escarpments or borders of forested areas, which cannot be correctly depicted in coarser resolutions (Fig.~\ref{fig:shade_high_slope}). 

Slope maps enhance local elevation differences, therefore are useful to visually assess the quality of DEMs. In all study areas, TanDEM-X data (12~m and 30~m) show no evidence of artificial artifacts introduced in the production process. SRTM presents striping artifacts in the flat areas of Arac\'a, Barcelos and Pantanal. These stripes are a well-known artifact \citep{Albani2003,Miliaresis2005,Gallant2009,Tarekegn2013,Crippen2016} and were caused by uncompensated oscillations in the SRTM mast that affected the interferometric baseline roll angle \citep{Farr2007,Simard2016}.

ASTER GDEM slope reveals issues such as noise in the Arac\'a, Barcelos and Rio Claro areas, as well as in the coastal plain of Santa Catarina. In the areas with more rugged topography, the noise is not so evident. Artifacts inherent to the automatic processing of optical imagery \citep{Reuter2009c,Miliaresis2011,Grohmann2015poznan} can be easily seen in the Arac\'a and Barcelos areas.

Although ALOS AW3D30 slope shows a similar behavior as the other datasets in terms of distribution of slope values per elevation and histogram characteristics, it is necessary to note the occurrence of large void areas in the AW3D30 data, particularly in the Arac\'a, Barcelos and Iporanga areas, likely caused by the dense cloud coverage present in these areas over the year.\\

\subsection{Contour lines}
\label{sec:contours}

\begin{figure*}[!hbt]
    \centering
    \includegraphics[width=0.9\textwidth]{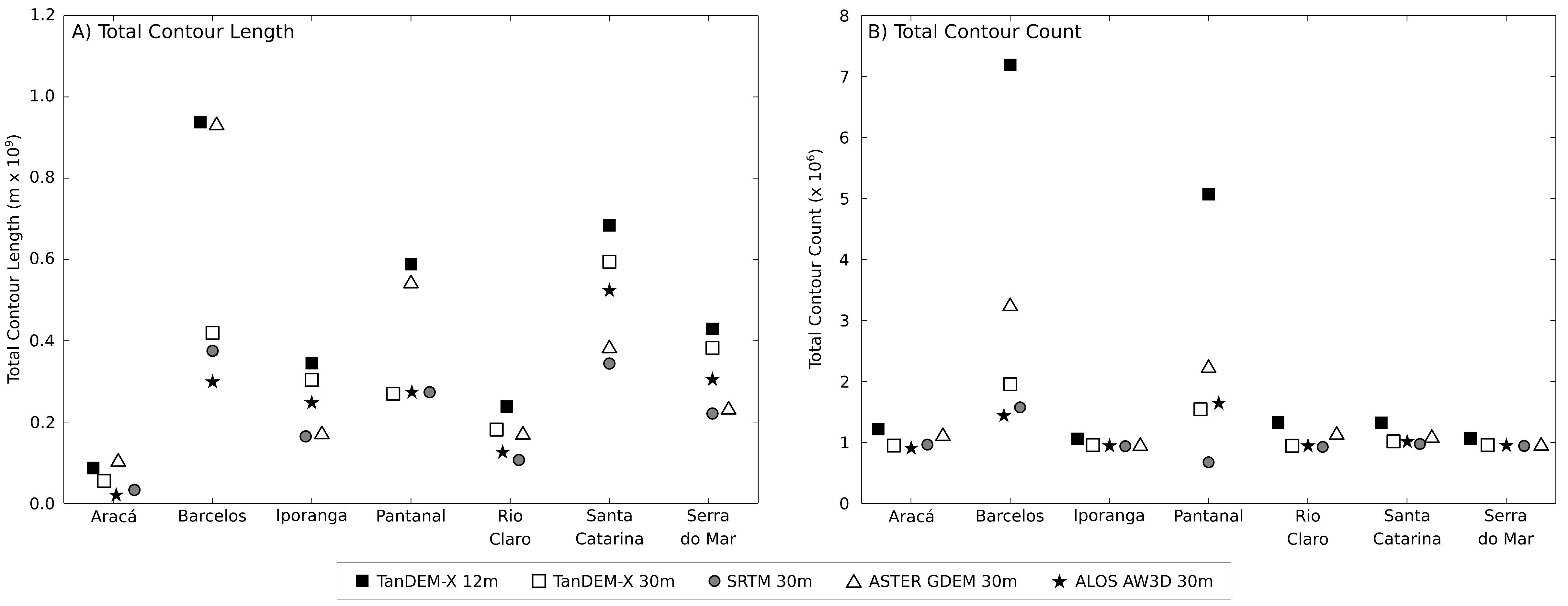} 
    \caption{A) Cumulative sum of contour line length. B) Cumulative sum of contour count for each DEM and study area.}
    \label{fig:cumsum_contours}
\end{figure*}

Contour lines are easily derived from DEMs and can be used to evaluate quantitatively how different DEMs represent the same landforms. Similar DEMs would result in similar contours, both in terms of number and length of generated lines. A noisy DEMs will result in higher number and total length of contour lines, while a smoothed DEM can produce the same number of contours (depending on which levels are selected to be calculated), but the length of lines will tend to be shorter. 

Contours were generated for each DEM at its original resolution (12~m or 30~m), with vertical interval according to the study area: Arac\'a: 20~m; Barcelos: 2~m; Pantanal: 2~m, Iporanga: 20~m; Rio Claro: 10~m; Santa Catarina: 5~m and Serra do Mar: 20~m. 

Statistics of generated contour lines are in Supplemental Table S15, plots of the cumulative sum of contour length in Supplemental Figure S19 and plots of cumulative sum of number of contours in Supplemental Figure S20. Selected examples of the contour lines derived from all DEMs are presented for Rio Claro, Serra do Mar and Iporanga areas in Supplemental Figures S21, S22, and S23. Figure~\ref{fig:cumsum_contours} shows summary plots of cumulative sum of contour line length and contour count for each DEM and study area.

Given its finer resolution, TanDEM-X 12~m shows higher total length as well as number of contours for all areas except Arac\'a, where the noise in ASTER GDEM results in higher total length.

ASTER GDEM has higher contour count than the other 30~m DEMs in all areas, but contour length is more dependent on landscape: in gentle topography areas (Arac\'a, Barcelos and Pantanal), contour length can be substantially higher than for other DEMs; in areas with steep relief (Iporanga, Santa Catarina and Serra do Mar), the length is similar to contours derived from SRTM, while in Rio Claro, with a landscape of rolling hills, contour length is similar to TanDEM-X 30~m.

TanDEM-X 30~m, SRTM and ALOS AW3D30 have similar contour count in all areas, with differences in contour length. In Arac\'a and Barcelos, ALOS AW3D30 has lower total length than TanDEM-X 30~m and SRTM due to large voids in the DEM. In the Pantanal, these 3 DEMs produce nearly identical results, while in the other areas TanDEM-X 30~m has higher value of total contour length, indicating its finer effective spatial resolution and more detailed representation of the terrain, and ALOS AW3D30 shows contour length higher than SRTM, possibly related to the level of detail captured by the 5~m original dataset (see Sec.\ref{sec:compare_dems}).

\subsection{Differences in elevation from TanDEM-X 12~m}\
\label{sec:diffs}

Metrics of the differences in elevation for all areas and datasets, considering TanDEM-X 12~m as the reference value, are presented in Table~\ref{tbl:error_metrics_dems}. Descriptive statistics of the differences are presented in Supplemental Tables S16--S22, maps of differences are shown in Supplemental Figures S24--S30, histograms in Supplemental Figures S31--S37, scatterplots of the difference from TanDEM-X 12~m \emph{versus} aspect for each DEM in Supplemental Figures S38--S44 and scatterplots of the elevation of each DEM \emph{versus} TanDEM-X 12~m in Supplemental Figures S45--S51. 

TanDEM-X 30~m shows the smaller deviations from TanDEM-X 12~m with LE90 ranging from 1.62~m to 11.75~m and RMSE between 0.84~m and 7.15~m. SRTM 30~m shows LE90 from 3.59~m to 13.54~m and RMSE between 3.11~m and 8.24~m. For ASTER GDEM, LE90 ranges from 6.88~m to 24.37~m and RMSE from 5.20~m to 15.17~m. In the case of ALOS AW3D30, LE90 ranges from 3.36~m to 14.41~m and RMSE from 2.75~m to 9.43~m.

Deviations from TanDEM-X 12~m reaching tens to hundreds of meters occur in all areas and for all DEMs, being considered outliers as they are above LE99 values (Supplemental Tables S16--S22). 

\begin{table*}[!hbt]
    \caption{Error metrics of DEMs compared to TanDEM-X 12~m. ME: Mean Error; RMSE: Root Mean Square Error; STDE: Standard Deviation of Error; LE90: Linear Error at 90\% confidence; LE95: Linear Error at 95\% confidence; LE99: Linear Error at 99.73\% ($3\sigma$) confidence.}
    \label{tbl:error_metrics_dems}
    \begin{center}
    \resizebox{0.9\textwidth}{!} {
    \begin{tabular}{llrrrrrrrr}
    \toprule
    TanDEM-X 12~m - DEM   &  Area        &    ME &  RMSE  &  STDE &   LE90  &  LE95  & LE99   \\
    \midrule         
    TanDEM-X 30~m     &  Arac\'a         &  0.00 &  0.98  &  0.98 &  1.62 &  1.93 &  2.95 \\
    {}                &  Barcelos        &  0.00 &  1.28  &  1.28 &  2.10 &  2.50 &  3.83 \\
    {}                &  Pantanal        &  0.00 &  1.54  &  1.54 &  2.54 &  3.03 &  4.63 \\
    {}                &  Iporanga        &  0.00 &  0.84  &  0.84 &  1.38 &  1.65 &  2.52 \\
    {}                &  Rio Claro       &  0.00 &  1.03  &  1.03 &  1.70 &  2.02 &  3.09 \\
    {}                &  Santa Catarina  &  0.08 &  5.94  &  5.94 &  9.77 & 11.64 & 17.81 \\
    {}                &  Serra do Mar    & -0.02 &  7.15  &  7.15 & 11.75 & 14.01 & 21.44 \\
    \midrule      
    SRTM 30~m         &  Arac\'a         & -1.22 &  3.57  &  3.35 &  5.51 &  6.57 & 10.05 \\
    {}                &  Barcelos        & -0.51 &  3.11  &  3.07 &  5.05 &  6.01 &  9.20 \\
    {}                &  Pantanal        & -2.32 &  6.75  &  6.34 & 10.43 & 12.43 & 19.02 \\
    {}                &  Iporanga        & -3.37 &  4.02  &  2.18 &  3.59 &  4.27 &  6.54 \\
    {}                &  Rio Claro       & -4.89 &  5.95  &  3.40 &  5.59 &  6.66 & 10.19 \\
    {}                &  Santa Catarina  & -0.52 &  7.03  &  7.01 & 11.54 & 13.75 & 21.04 \\
    {}                &  Serra do Mar    & -0.26 &  8.24  &  8.23 & 13.54 & 16.13 & 24.70 \\
    \midrule  
    ASTER GDEM 30~m   &  Arac\'a         &  3.28 & 15.17  & 14.81 & 24.37 & 29.03 & 44.44 \\
    {}                &  Barcelos        & -1.37 &  8.72  &  8.61 & 14.16 & 16.88 & 25.83 \\
    {}                &  Pantanal        &  1.60 &  9.20  &  9.06 & 14.90 & 17.76 & 27.18 \\
    {}                &  Iporanga        & -3.10 &  5.20  &  4.18 &  6.88 &  8.20 & 12.55 \\
    {}                &  Rio Claro       & -3.74 &  8.19  &  7.28 & 11.98 & 14.27 & 21.84 \\
    {}                &  Santa Catarina  &  1.48 & 11.06  & 10.96 & 18.02 & 21.48 & 32.87 \\
    {}                &  Serra do Mar    &  4.66 & 14.83  & 14.08 & 23.16 & 27.60 & 42.25 \\
    \midrule      
    ALOS AW3D30 30~m  &  Arac\'a         & -1.02 &  2.79  &  2.60 &  4.28 &  5.10 &  7.80 \\
    {}                &  Barcelos        & -0.59 &  2.75  &  2.69 &  4.42 &  5.27 &  8.06 \\
    {}                &  Pantanal        & -3.48 &  9.43  &  8.76 & 14.41 & 17.17 & 26.29 \\
    {}                &  Iporanga        & -3.56 &  4.11  &  2.04 &  3.36 &  4.00 &  6.12 \\
    {}                &  Rio Claro       & -4.85 &  5.74  &  3.06 &  5.03 &  5.99 &  9.17 \\
    {}                &  Santa Catarina  & -1.95 &  5.41  &  5.04 &  8.30 &  9.89 & 15.13 \\
    {}                &  Serra do Mar    & -1.59 &  5.36  &  5.11 &  8.41 & 10.02 & 15.34 \\
    \bottomrule
    \end{tabular}
    } 
    \end{center}
\end{table*}

DEMs of differences (DoDs) are useful to highlight issues inherent to the production methods of the analyzed DEMs, which are observed in all study areas. Mast oscillations in SRTM data are clearly visible when comparing SRTM with TanDEM-X data. The DoDs for ASTER GDEM shows strong noise and NNE-SSW oriented swaths, corresponding to overlap areas of the original images; a significant difference in elevation can be seen between adjacent swaths in some cases. In the DoDs for ALOS AW3D30, besides the large areas without information (voids), scene boundaries can also be seen, sometimes forming a crisscross pattern. 

In the Arac\'a area (Supplemental Figure S24), the larger errors are localized in the steep slopes of the tepuis, which highlight the problem of reconstructing the topography of this kind of landform due the geometry of the terrain relative to the radar sensor (SRTM), or with photogrammetry (ASTER, ALOS), since some of these areas tend to be covered by shadows at the time of image acquisition. 

\begin{figure*}[!hbt]
    \centering
    \includegraphics[width=0.85\textwidth]{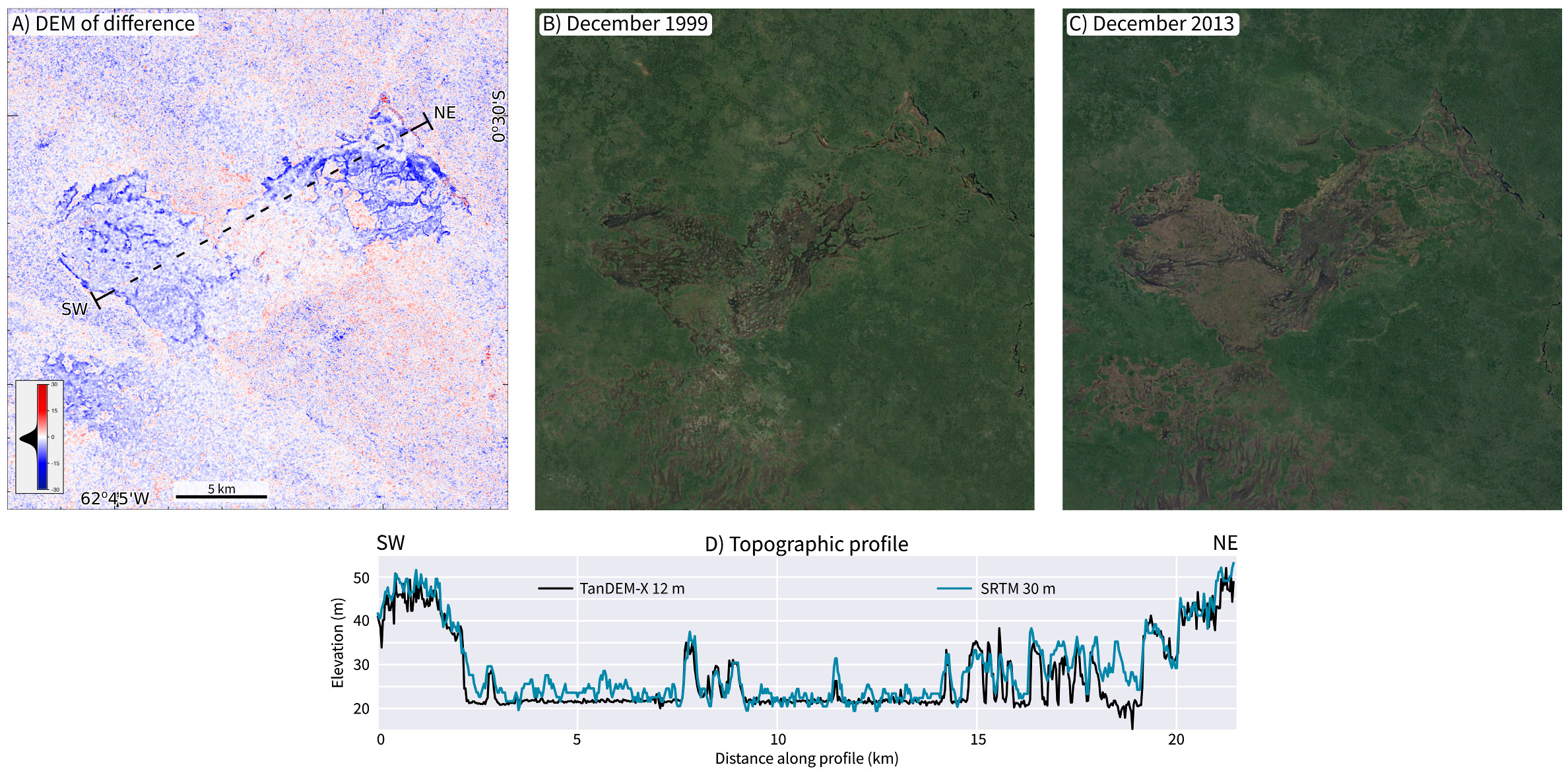} 
    \caption{Differences between SRTM and TanDEM-X as a result of natural changes in the environment (see location in Fig.\ref{fig:areas}B). A) Dem of difference (TanDEM-X 12~m $-$ SRTM) for the Barcelos area, with indication of topographic profile location; B) Landsat image of 12-30-1999; C) Landsat image of 12-30-2013; D)Topographic profile; Satellite imagery \textcopyright Landsat/Copernicus, powered by Google.}
    \label{fig:diff_barcelos}
\end{figure*}

The DoD for SRTM of the Barcelos area (Supplemental Figure S25) shows irregular shapes in its central-western region, where SRTM is higher than TanDEM-X, which correspond to wetlands with natural (i.e, non-anthropogenic) changes in its vegetation cover from 2000 (SRTM acquisition) to 2013 (TanDEM-X acquisition). These changes can be observed in Fig.~\ref{fig:diff_barcelos}, where the satellite image of 2013 (Fig.~\ref{fig:diff_barcelos}C) shows less vegetation in the wetlands area than in 1999 (Fig.~\ref{fig:diff_barcelos}B).

Elevation of SRTM at the Pantanal area (Supplemental Figure S26) is generally higher than TanDEM-X, likely related to the finer effective spatial resolution of TanDEM-X, capable of penetrating the open vegetation. In the southeast quadrant of the area, two patches in darker blue are related to deforestation between 2000 and 2013.

\begin{figure*}[!hbt]
    \centering
    \includegraphics[width=0.85\textwidth]{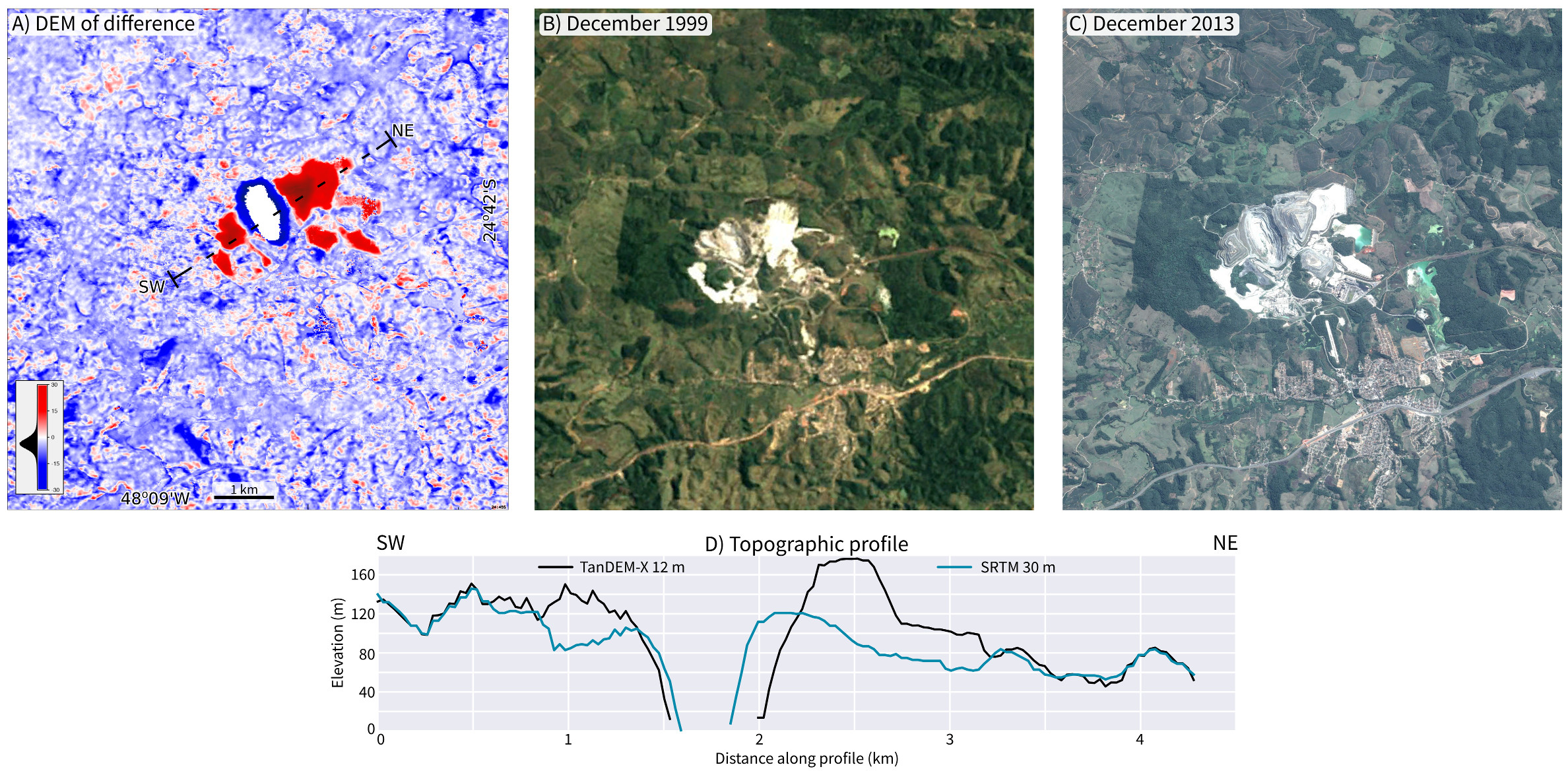} 
    \caption{Differences between SRTM and TanDEM-X caused by mining activity (see location in Fig.\ref{fig:areas}D). A) Dem of difference (TanDEM-X 12~m $-$ SRTM) for the Iporanga area, with indication of topographic profile location; B) Landsat image of 12-30-1999; C) Landsat image of 12-30-2013; D)Topographic profile; Satellite imagery \textcopyright Landsat/Copernicus, powered by Google.}
    \label{fig:diff_iporanga}
\end{figure*}

Several patches of red and blue are observed in the northern sector of the Iporanga area (Supplemental Figure S27). These patches are located mainly in a plateau of smooth topography and represent agricultural lands where crops were mature at the time of each DEM's acquisition. In the east-southeast sector, anthropic change in the landscape is characterized by a mining activity. In Figure \ref{fig:diff_iporanga}, the expansion of the quarry and accumulation of residual material is visible in the satellite images and in the topographic profile.

In the Rio Claro area (Supplemental Figure S28), SRTM elevation is moderately higher than TanDEM-X. Localized larger differences occur in agricultural lands. Urban growth was not detected in the DoDs, even tough it can be observed from the archive of historical orbital imagery available in virtual globe applications such as Google Earth (\url{https://www.google.com/earth/}) or NASA's WorldWind (\url{https://worldwind.arc.nasa.gov}).

The coastal areas of Santa Catarina and Serra do Mar (Supplemental Figures S29 and S30) show similar results for the DoDs of TanDEM-X 30~m and SRTM. Localized patches of significant differences in the SRTM DoD are related to deforestation and agriculture. In both areas it is possible to observe that TanDEM-X 12~m is higher (red) than TanDEM-X 30~m and SRTM in E-SE slopes but lower (blue) in N-NW slopes of local ranges. This effect is not observed in the DoDs for ALOS AW3D30 in the same areas. 

\begin{figure*}[!hbt]
    \centering
    \includegraphics[width=0.85\textwidth]{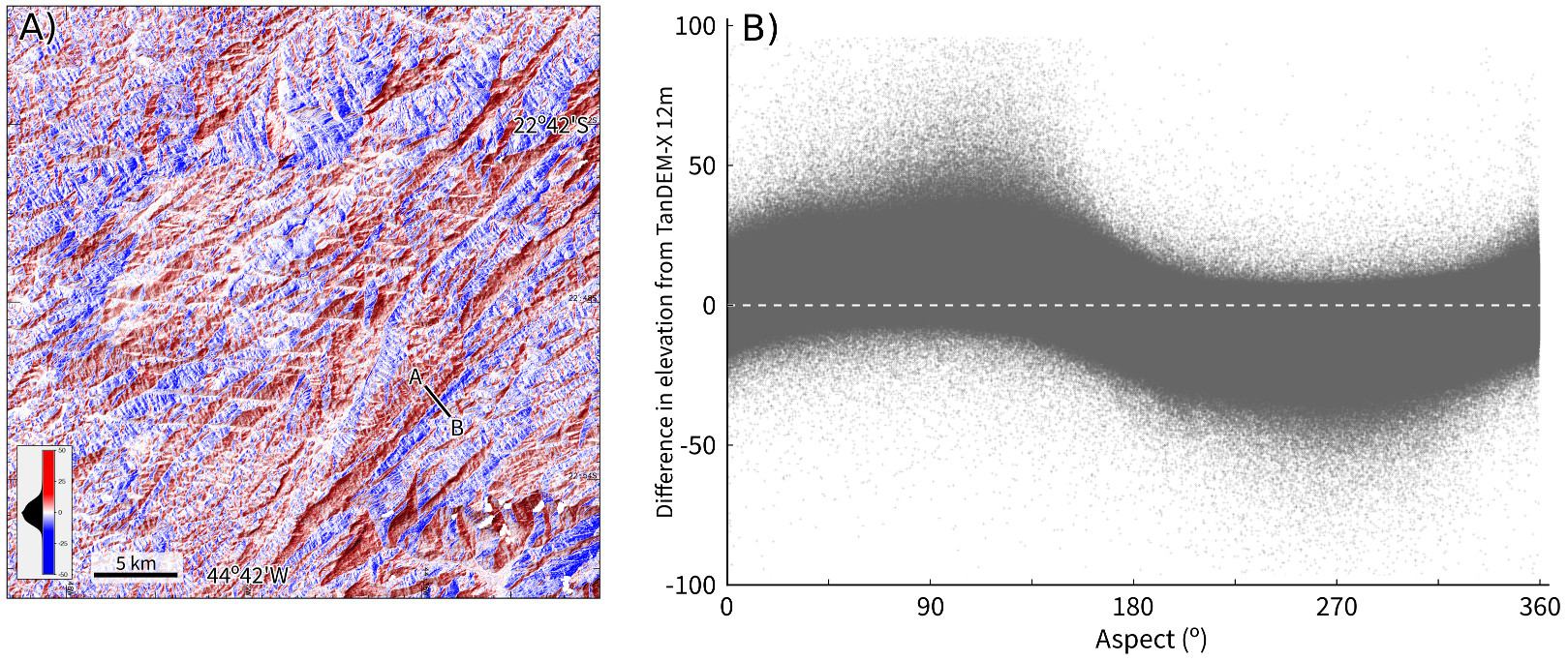} 
    \caption{A) DoD for TanDEM-X 12~m minus TanDEM-X 30~m; B) scatterplot of TanDEM-X 30~m aspect \emph{versus} deviations from TanDEM-X 12~m.}
    \label{fig:dod_aspect}
\end{figure*}

Figure \ref{fig:dod_aspect} shows, for a subset of the Serra do Mar area (see Fig.~\ref{fig:areas}G for location), the DoD and a scatterplot of aspect of TanDEM-X 30~m \emph{versus} deviations from TanDEM-X 12~m. There is a clear tendency of positive differences (where TanDEM-X 30~m have lower elevations than TanDEM-X 12~m) for `eastern' aspects ($\sim\ang{330}\rightarrow\ang{0}\rightarrow\ang{90}\rightarrow\ang{150}$) and negative differences for `western' aspects ($\sim\ang{150}\rightarrow\ang{180}\rightarrow\ang{270}\rightarrow\ang{330}$).

This could, at first, be attributed to a misregistration between the DEMs, although this wouldn't be the case of TanDEM-X 30~m since it is derived from TanDEM-X 12~m, and it would need to be observed in all study areas, but it is evident only in Santa Catarina, Serra do Mar and, to a lesser extent, Iporanga.

\begin{figure*}[!hbt]
    \centering
    \includegraphics[width=0.85\textwidth]{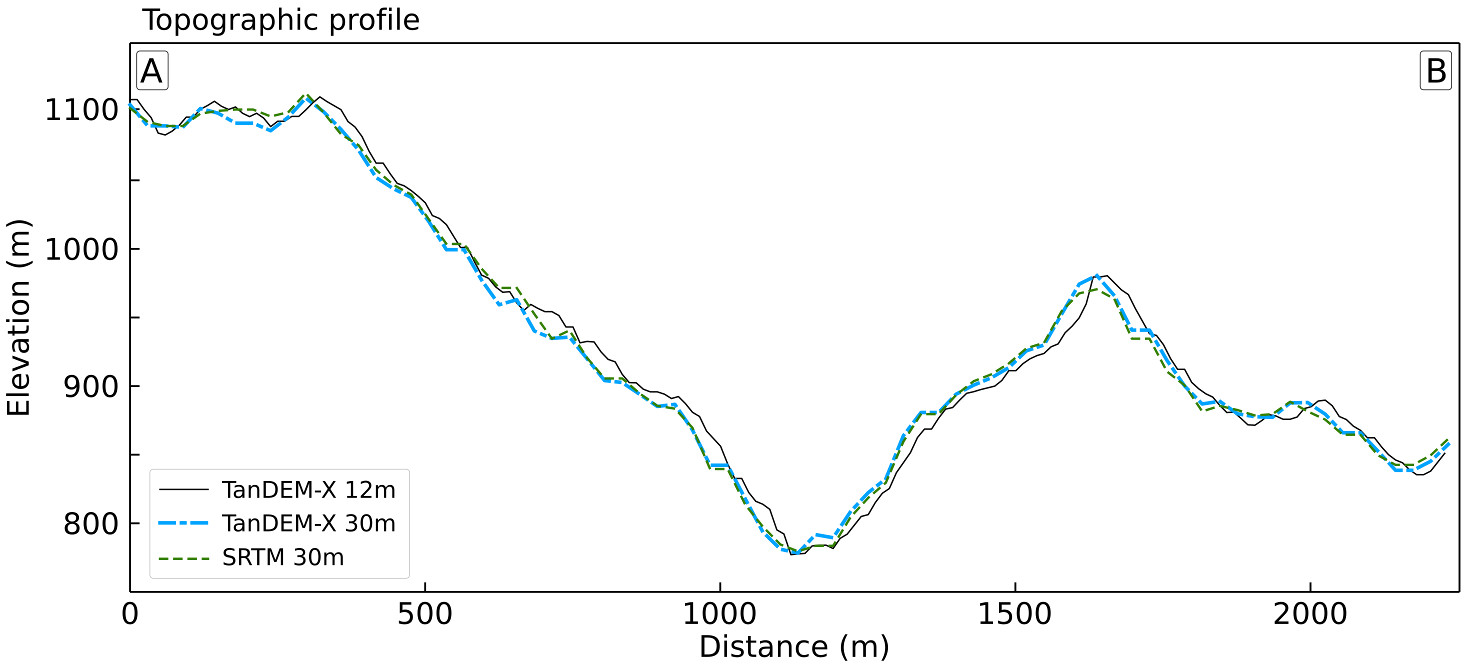} 
    \caption{Topographic profiles of TanDEM-X 12~m, TanDEM-X 30~m and SRTM 30~m. Location of profile on Fig.~\ref{fig:dod_aspect}A.}
    \label{fig:profile_mw}
\end{figure*}

The topographic profile in Figure~\ref{fig:profile_mw} shows that TanDEM-X 30~m and SRTM plot slightly above or below TanDEM-X 12~m depending on the configuration of the landscape. This is caused by the moving-window process used to resample the data to a 30~m pixel size. 

Moving-window is a common method to filter or smooth raster data based on the values within a window or neighborhood. The new value for a pixel is determined by a function (such as mean or standard deviation) of all pixel values inside an $n\times n$ window (usually with odd $n$) centered in the pixel of interest; the window moves one pixel at a time across the raster map, until the whole area is processed \citep{DeMers2004,Lillesand2004,Grohmann2009cageo}. Since the origin of raster maps (row $J=0$, column $I=0$) is the upper left corner with increasing $I$ to the right and $J$ down \citep[e.g.,][]{Ritter1997}, a moving-window operation will run from west to east and from north to south. 

With slopes facing south, pixel values will be progressively lower `ahead' of the moving-window and their arithmetic mean will tend to be lower than the original central pixel value, resulting in an area of the DEM with elevations below the original ones. Conversely, slopes facing north will resulting in higher elevations in that area than in the original DEM.

One possible solution to this unintended result is to produce a 30~m DEM based on the original value of the central pixel in the moving-window, similar to SRTM version 2 \citep{Farr2007}, although the 12~m original resolution does not favor resampling to 30~m. A 25~m pixel would be more suitable and could be produced by averaging pixel values in a $2\times 2$ window.

\section{Conclusions}
\label{sec:conclusions}

A first assessment of the TanDEM-X DEMs over Brazilian territory was presented through a comparison with SRTM, ASTER GDEM and ALOS AW3D30 DEMs in seven study areas with distinct geomorphological contexts, vegetation coverage and land use. 

A correct application of the Water Indication Mask (WAM), supplied as an auxiliary file of TanDEM-X data, is essential to remove areas with random noise that can influence the results of morphometric analysis, although it must be done carefully to avoid masking areas which were flagged for amplitude or coherence but don't represent water bodies.

ASTER GDEM showed a noisy surface and strong differences in elevation between adjacent scenes. In open areas ALOS AW3D30 can be comparable to TanDEM-X, visually and about the representation of terrain by contour lines, but it also shows mismatch between adjacent scenes and has large areas of voids, caused by cloud coverage in the original imagery. SRTM has been extensively used, and provides a good representation of the topography, although its effective spatial resolution is coarser than the nominal 30~m. 

DEMs of differences (DoDs) allowed the identification of issues inherent to the production methods of the analyzed DEMs, such as mast oscillations in SRTM data and scene mismatch in ASTER GDEM and ALOS AW3D30. It is strongly recommended to produce a DoD with SRTM before using ASTER GDEM or ALOS AW3D30 in any analysis, to evaluate if the area of interest is affected by these problems.

A systematic difference in elevation was observed in the steep slopes of Serra do Mar, Santa Catarina and Iporanga, where TanDEM-X 12~m has higher elevations than TanDEM-X 30~m and SRTM in E-SE slopes but lower elevations in N-NW slopes. Related to the moving-window process used to resample the 12~m data to a 30~m pixel size, it could be solved by producing a DEM with 25~m resolution by averaging pixel values in a $2\times 2$ window.

TanDEM-X was built to represent a new standard in global DEMs, with remarkable level of detail and consistency, although its commercial distribution might hinder a wide adoption by researchers in the immediate future. Further evaluations of this dataset should involve comparisons with other sources of elevation data including local LiDAR surface/terrain models and global DEMs such as the MERIT DEM and the upcoming NASADEM, as well as geomorphometric analyses, landslide characterization and hydrological modeling. A global forest/non-forest map derived from TanDEM-X data has been recently presented by \cite{Martone2018}; the dataset is expected to be released freely to the scientific community and can be of great value in assessing the influence of land cover type, such as open vegetation (savanna), in the representation of the topographic surface by InSAR DEMs.

\section*{Acknowledgements}

This study was supported by Brazil's National Council of Scientific and Technological Development, CNPq grant 307647/2015-3, the Sao Paulo Research Foundation (FAPESP) grant \#2016/06628-0, and is co-funded by FAPESP (BIOTA \#2012/50260-6, \#2013/50297-0), NSF (DEB 1343578), and NASA. Acknowledgements are extended to the Editor-in-Chief and the anonymous reviewers for their criticism and suggestions, which helped to improve this paper. TanDEM-X data was provided by the German Aerospace Centre (DLR) through an Announcement of Opportunity \& Proposal Call (proposal DEM\_GEOL0538). TanDEM-X data \textcopyright DLR 2017. SRTM data (SRTMGL1-V003) courtesy of the NASA EOSDIS Land Processes Distributed Active Archive Center (LP DAAC), USGS/Earth Resources Observation and Science (EROS) Center, Sioux Falls, South Dakota. ASTER GDEM is a product of METI and NASA. ALOS AW3D30 data \textcopyright JAXA.

\appendix

\section{Supplementary material}
Supplementary material associated with this article can be found at GitHub: \url{https://git.io/vQTyp}.

\section*{References}
\label{references}


\end{document}